\documentclass[smallextended]{svjour3}
\smartqed 
\usepackage{graphicx}
\usepackage{epstopdf}
\usepackage{setspace}
\usepackage{epsf}
\usepackage{mathtools}

\begin{document}

\title{Lumps and Rogue waves of Generalized Nizhnik Novikov Veselov  Equation} 
\author{P. Albares \and P. G. Estevez \and R. Radha \and R. Saranya 
}

\institute{P. Albares \and  P. G. Estevez \at
              Departamento de F\'isica Fundamental, Universidad de Salamanca, Salamanca\textendash{}E-37008, Spain. \\
              \email{pilar@usal.es}         
           \and
          R. Radha \and R. Saranya \at
             Centre for Nonlinear Science (CeNSc), Post-Graduate and Research Department
of Physics, Government College for Women (Autonomous), Kumbakonam\textendash{}612001, India. \\
\email{vittal.cnls@gmail.com}  
      \and
      R. Saranya \at
      Post-Graduate and Research Department of Mathematics, Government College for Women (Autonomous), Kumbakonam\textendash{}612001, India.\\
}
\date{Received: date / Accepted: date}
\maketitle

\begin{abstract}
We investigate the generalized $(2+1)$ Nizhnik-Novikov-Veselov equation and construct its linear eigenvalue problem in the coordinate space from the results of singularity structure analysis thereby dispelling the notion of weak Lax pair. We then exploit the Lax-pair employing Darboux transformation and generate lumps and rogue waves. The dynamics of lumps and rogue waves is then investigated.\\
PACS: 02.30 lk, 05.45 Yv, 02.30 Jr
\keywords{Lumps \and Rogue waves \and Singular manifold method \and Partial Differential Equations}
\end{abstract}

\section{Introduction}
The identification of dromions \cite{one,fokas} in the Davey-Stewartson I (DSI) equation which has given a fillip to the investigation of $(2+1)$ dimensional integrable nonlinear partial 
differential equations (pdes) \cite{Chen} has virtually triggered
a renewed interest towards other localized structures like lumps \cite{estevez1}, breathers \cite{ling} etc. Recent identification of rogue waves \cite{chang,Wen} in nonlinear pdes which appear from nowhere has once again prompted a deeper investigation of integrable $(2+1)$ nonlinear pdes in an effort to unearth similar structures in them. It should also be mentioned that even though  the integrability of $(2+1)$ dimensional nonlinear pdes has been well established in terms of the abundance of localized solutions, there exists no systematic approach to unearth other signatures of integrability like Lax pair \cite{estevez1}, B\"acklund transformation 
\cite{Backlund}, Hamiltonian Structures \cite{huan}, conservation laws 
\cite{xia} etc.. In this connection, Boiti et al. \cite{boiti1,boiti2} had pointed out that $(2+1)$ dimensional nonlinear pdes like Nizhnik- Novikov -Veselov (NNV) equation \cite{Rogers} admits only weak Lax pair in the subspace of coordinate space. In other words, the lax operators commute at least on  the functional subspace of the eigenfunction and they should be compatible at least
 for one eigenvalue. Even though the concept of weak lax pair has yielded several $(2+1)$ integrable nonlinear pdes and facilitated
their investigation from the viewpoint of localized coherent structures ~\cite{radha1,radha2}, a closer look at the investigation of integrable
 $(2+1)$ nonlinear pdes may yield other richer structures and would enable us to get a deeper understanding of integrability.
 
 The Painlev\'e property \cite{painleve} has been proved to be a powerful test for identifying the integrability as well as a good basis for the determination of many of the properties derived of the integrability of a given pde \cite{estevez1}.
  In this paper, we investigate the $(2+1)$ dimensional generalized Nizhnik- Novikov- Veselov equation \cite{radha1} and generate the Lax pair in the coordinate  space employing the singular manifold method \cite{weiss} based on the Painlev\'e analysis.
 We then exploit the Lax pair employing Darboux transformation approach, and construct lumps and rogue waves. We then discuss their dynamics.
 
The present paper is structured as follows: in section 2, we drive the linear eigenvalue problem of the NNV equation by using the results of Painlev\'e analysis. We then exploit the Lax pair and employ Darboux transformation in section 3, to derive lumps in section 4 and rogue waves in section 5. After studying the dynamics of lumps and rogue waves, the results are summarized at the end.

\section{Singular Manifold Method for the Nizhnik-Novikov-Veselov equation}
The generalized  Nizhnik-Novikov-Veselov (NNV)
 equation is a symmentric generalization of the KdV equation to $(2+1)$ dimensions
and is given by
\begin{eqnarray}
u_t + au_{xxx}+bu_{yyy}+c~u_x+d~u_y-3~a(uv)_x
-3~b (uw)_y = 0 \label{1a}\\
u_x = v_y \label{1b}\\
u_y=w_x \label{1c}
\end{eqnarray}
where $a,b, c$ and $d$ are parameters. This equation, which is also known to be completely integrable, has been investigated in \cite{radha1,radha2} where exponentially localized solutions have been generated and their dynamics has been investigated.
Introducing the  following change of variables,
\begin{eqnarray}
u= -2~ m_{xy},~~
v= \frac{c}{3~a} -2~ m_{xx} ,~~
w =\frac{d}{3 ~b} - 2~ m_{yy} \label{2a}
\end{eqnarray}
Eqs.~(\ref{1a})-(\ref{1c}) get converted to the following equation:
\begin{eqnarray}
m_{xyt} +a~(m_{xxxy} +6~ m_{xx} m_{xy})_x +b~(m_{yyyx}
+6 ~m_{yy} m_{xy})_y =0 \label{3}
\end{eqnarray}
According to the  singular manifold method \cite{weiss,estevez2},  the truncated Painlev\'e expansion
 for $m$ should be
\begin{eqnarray}
m^{[1]} = m^{[0]} + \ln(\phi_1) \label{4}
\end{eqnarray}
    where $m^{[1]}$ and $m^{[0]}$ are both solutions of Eq.~(\ref{3}) and $\phi_1$ is the singular manifold for the seed solution $m^{[0]}$. Furthermore, Eq.~(\ref{4})  also implies an iterative method of constructing solutions where the super index $[0]$ denotes a seed solution and $[1]$ the iterated one.
Substitution of Eq.~(\ref{4})  into Eq.~(\ref{3}) yields an expression in negatives powers of $\phi_1$.
Eq.~(\ref{3}) is symmetric under the interchange of $(x,a)$ and $(y,b)$ and hence it is  reasonable to suggest the ansatz,
\begin{eqnarray}
\phi_{1,t} = a~G_a(x,y,t) +b~ G_b(x,y,t)\label{6}
\end{eqnarray}
such that the terms in $a$ and $b$ cancel independently. Substituting equation Eq.~(\ref{6})  into the expression in negatives powers of $\phi_1$, we obtain two polynomials(one for the terms in $a$ and other for the terms in $b$) in negative powers  of $\phi_1$. If we require all the coefficients of these polynomials to be zero, we obtain the following  expressions after some algebraic  manipulations [using Maple]. The result can be summarized as follows:
\begin{eqnarray}
G_a = -\phi_{1,xxx} - 6 ~\phi_{1,x} m^{[0]}_{xx},
~~G_b = -\phi_{1,yyy} - 6~ \phi_{1,y} m^{[0]}_{yy} \label{7b}
\end{eqnarray}
The rest of the terms can be independently integrated as,
\begin{eqnarray}
\frac{\phi_{1,xy} +2 ~\phi_1 m^{[0]}_{xy} }{\phi_{1,x} }  + K_2(y) + K_1(y)  \int \left(\frac{\phi_1}{\phi_{1,x}}\right)^2 dx = 0 \label{8a}\\
\frac{\phi_{1,xy} +2 ~\phi_1 m^{[0]}_{xy} }{\phi_{1,y} }  + H_2(x) + H_1(x) \int \left(\frac{\phi_1}{\phi_{1,y}}\right)^2 dy = 0 \label{8b}
\end{eqnarray}
where $H_i (x)$ and $K_i (y)$ are arbitrary functions.
Comparison of Eqs.~(\ref{8a})-(\ref{8b}) yields ( with $H_1 = H_2= K_1 = K_2 = 0$) and therefore,
\begin{eqnarray}
\phi_{1,xy} +2~ \phi_1 m^{[0]}_{xy} =0 \label{9}
\end{eqnarray}
and the combination of Eq.~(\ref{6})  and  Eq.~(\ref{7b}) yields,
\begin{eqnarray}
\phi_{1,t} + a ~(\phi_{1,xxx} + 6 ~\phi_{1,x} m^{[0]}_{xx}) + b~ (\phi_{1,yyy}
+ 6~ \phi_{1,y} m^{[0]}_{yy})=0\label{10}
\end{eqnarray}
Eq.~(\ref{9}) and Eq.~(\ref{10}) constitute the Lax pair for the  NNV equation (\ref{3}).
 The above Lax pair is in sharp contrast to the notion of weak Lax pair postulated by Boiti et al.\cite{boiti1,boiti2} in the subspace of coordinate space.
\section{Darboux Transformations} 
The truncated expansion given by Eq.~(\ref{4}) can be considered as an iterative method   \cite{estevez1,estevez2} such that an iterated solution $m^{[1]}$ can be obtained from the seed solution $m^{[0]}$, if we know a solution $\phi_1$ for the Lax pair of this seed solution. This means that if we denote $\phi_{1,2}$ as the eigenfunction for the iterated solution $m^{[1]}$, it should satisfy the following Lax pair,
\begin{eqnarray}
&&(\phi_{1,2})_{xy} +2 ~\phi_{1,2} m^{[1]}_{xy} =0 \label{11a}\\
\nonumber
&&(\phi_{1,2})_{t}+a \left[(\phi_{1,2})_{xxx} +6  (\phi_{1,2})_{x} m^{[1]}_{xx}\right]
 +b \left[(\phi_{1,2})_{yyy} +6 (\phi_{1,2})_{y} m^{[1]}_{yy}\right]=0  \label{11b}\\
\end{eqnarray}
The Lax pair can also be considered as a nonlinear system  between the fields and eigenfunction together \cite{estevez1,estevez2}. It means that the truncated Painlev\'e expansion given by Eq.~(\ref{4}) should be combined in Eqs.~(\ref{11a})-(\ref{11b}) with a similar expansion for the  eigenfunction such as,
\begin{eqnarray}
\phi_{1,2} = \phi_2 -\frac{\Delta_{1,2}}{\phi_1} \label{13}
\end{eqnarray}
where $\phi_i$, (i=1,2) are eigenfunctions for the seed solution $m^{[0]}$ and therefore,
\begin{eqnarray}
&&\phi_{i,xy} +2~ \phi_i m^{[0]}_{xy} =0 \label{14a}\\
&&\phi_{i,t}+a~\left(\phi_{i,xxx} +6 ~\phi_{i,x} m^{[0]}_{xx}\right)
+b~\left(\phi_{i,yyy} +6~ \phi_{i,y} m^{[0]}_{yy}\right)=0 \label{14b}
\end{eqnarray}
Substitution of Eqs.~(\ref{4}) and (\ref{13}) into Eqs.~(\ref{11a})-(\ref{11b}) yields  $\Delta_{i,j}$ as the exact derivative
\begin{eqnarray}
\nonumber
d\Delta_{i,j} = 2~ \phi_j ~\phi_{i,x} dx + 2 ~ \phi_{j,y} \phi_{i} dy\\
\nonumber
+2 a \left(\phi_{j,x} \phi_{i,xx} - \phi_{i,x} \phi_{j,xx} -\phi_j \phi_{i,xxx}
-6 ~m^{[0]}_{xx} \phi_j \phi_{i,x}\right) dt \\
+2~ b~ \left(\phi_{i,y} \phi_{j,yy}-\phi_{j,y} \phi_{i,yy}
 -\phi_i \phi_{j,yyy}
 -6 ~m^{[0]}_{yy} \phi_{i} \phi_{j,y}\right) dt \label{15}
\end{eqnarray}
where
\begin{eqnarray}
\Delta_{i,j} =2 ~\phi_i \phi_j -\Delta_{j,i} \label{16}
\end{eqnarray}
The Painlev\'e expansion given by Eq.~(\ref{4}) and Eq.~(\ref{13}) can be also considered as a binary  Darboux transformation that relates the Lax pairs given by Eqs.~(\ref{11a})-(\ref{11b}) and Eqs.~(\ref{14a})-(\ref{14b}).
\subsection{Iterated solution}
 In the previous section, we have introduced a singular manifold $\phi_{1,2}$ which allows us to iterate Eq.~(\ref{4}) again in the following form:
 $m^{[2]} = m^{[1]} +\ln(\phi_{1,2}) = m^{[0]} +\ln(\tau_{1,2})$,
 where $\tau_{1,2}$ is the $\tau$- function defined as,
\begin{eqnarray}
\tau_{1,2} =\phi_{1,2}~ \phi_1 =\phi_1~ \phi_2 - \Delta_{1,2} \label{18}
\end{eqnarray}
From Eq.~(\ref{16}), $\Delta _{1,2} = 2 ~\phi_1~ \phi_2 - \Delta_{2,1}$. If $\tau^2_{12} = det(\Delta_{i,j})$
~where $i,j = 1,2$. Therefore, we can construct the solution $m^{[2]}$ for the second iteration with just the knowledge of two eigenfunctions $\phi_1$ and $\phi_2$ for the seed solution $m^{[0]}$.
\section{Lumps}
In this section, we obtain lumps for the generalized NNV Eq.~(\ref{3}).
\subsection{Seed solution and Eigenfunction}
We consider a seed solution of the form,
\begin{eqnarray}
m^{[0]} =q_0 xy \label{20}
\end{eqnarray}
where $q_0$ is an arbitrary constant. Solutions of Eqs.~(\ref{14a})-(\ref{14b}) can be obtained  through the following form,
\begin{eqnarray}
&&\phi_i(k_i) = exp^{k_i(x+J(k_i))} P^n(k_i) \label{21a},~~
J(k_i) = -2 \frac{q_0 }{k_i^2} y +\left(-a k^2_i +\frac{8bq_0^3}{k_i^4}\right) t  \label{21b}\\
\nonumber
~~
&&P^n(k_i) = \sum^{n}_{j=0} a_j(k_i) \psi(k_i)^j \label{21c},~~
\psi(k_i) = k_i^2 \left( x+ \frac{2 q_0 }{k_i^2} y -3\left( a k_i^2 +\frac{8bq_0^3}{k_i^4}\right) t\right) \label{21d}\\
\end{eqnarray}
such that  $P^n(k_i)$ is a polynomial in $x$  of degree $n$ whose coefficients  $a_i$ can be obtained by substituting Eqs.~(\ref{21a})-(\ref{21d}) into Eqs.~(\ref{14a})-(\ref{14b}). We obtain after some algebraic calculation,
\begin{eqnarray} 
&&\frac{\partial a_j}{\partial y} = -k_i (j+1) \frac{\partial a_{j+1}}{\partial y} -2 q_0 k_i (j+1)(j+2) a_{j+2} \label{22a}\\
\nonumber
&&\frac{\partial a_j}{\partial t} = (j+1) \left( b k_i \frac{\partial^3 a_{j+1}}{\partial y^3} -12 b q_0 \frac{\partial^2 a_{j+1}}{\partial y^2}
 + \frac{36 b q^2_0}{k_i}
 \frac{ \partial a_{j+1}}{\partial y}\right)
+ (j+1)(j+2)\\ 
\nonumber
&&\left( 2 b q_0 k_i \frac{\partial^2 a_{j+2}}{\partial y^2}
- 24 b q_0^2 \frac{\partial a_{j+2}}{\partial y}-\frac{3}{k_i}\left(ak_i^6-16bq_0^3\right)a_{j+2}\right)\\
&&-(j+1)(j+2)(j+3)(a k_i^6 + 8 b q_0^3) a_{j+3} \label{22b}
\end{eqnarray}
where we can set $a_n=1$, $a_{n-1}=0$. From the above, it is obvious that there are an infinite number of possible  eigenfunctions  characterized by an integer $n$ and a wave number $k_i$.
\subsection{Case-I: $n=1$}
The simplest case can be obtained by taking $n=1$, in which case the eigenfunction given by Eqs.~(\ref{21a})-(\ref{21d}) is of the following form,
\begin{eqnarray}
\phi_{i}(k_{i}) = exp^{k_{i} (x+J(k_{i}))} \psi(k_{i}),~
J(k_{i}) = \frac{-2 q_0 y}{k^2_{i}} +\left(-a k^2_{i} +\frac{8 b q_0^3}{k^4_{i}}\right) t \label{23a} \\
\psi(k_{i}) =k^2_{i}\left(x+\frac{2 q_0 y}{k^2_{i}} -3\left(a k^2_{i} +\frac{8 b q^3_0}{k^4_i}\right)t\right) \label{23b}
\end{eqnarray}
According to Eq.~(\ref{15}), we can calculate $\Delta_{i,j}$ as,
\begin{eqnarray}
\nonumber
\Delta_{i,j} =\frac{2 k_i}{k_i+k_j} \left[\left(\psi(k_i)+ \frac{k_i k_j}{k_i+k_j}\right) \left( \psi(k_j)- \frac{k^2_j}{k_i+k_j}\right)+\frac{k^2_i k^2_j}{(k_i+k_j)^2}\right]\\
exp^{k_i(x+J(k_i))+k_j(x+J(k_j))} \label{24}
\end{eqnarray}
It is important to note that, $2 \phi_i \phi_j =\Delta_{i,j}+ \Delta_{j,i}$
 \subsection*{$\tau - function$}
A second iteration provides,
\begin{eqnarray}
m^{[2]} = m^{[0]} +\ln(\tau_{1,2}) \label{25}
\end{eqnarray}
Substituting Eq.~(\ref{20}) into Eq.~(\ref{25}), we obtain,
\begin{eqnarray}
\nonumber
m^{[2]} = q_0 x y +\ln(\tau_{1,2})
\end{eqnarray}
From Eq.~(\ref{2a}), we have
\begin{eqnarray}
\nonumber
u^{[2]} =-2 m^{[2]}_{x,y}=
-2\left( q_0 +\left(\frac{(\tau_{1,2})_x}{\tau_{1,2}}\right)_y\right)
\end{eqnarray}
where ~$\tau_{1,2}  = \phi_1 \phi_2 -\Delta_{1,2} = \frac{1}{2} \left(\Delta_{2,1} -\Delta_{1,2}\right)$, which after simplification can be written as,
\begin{eqnarray}
\nonumber
\tau_{1,2} = - \frac{k_1 - k_2}{k_1 +k_2} exp^{k_{1} (x+J(k_{1}))+k_2(x+J(k_{2}))} \Omega_{1,2}\\
\Omega_{1,2} = (\psi(k_1)+g(k_1,k_2))(\psi(k_2)+g(k_2,k_1))+d(k_1,k_2) \label{26}
\end{eqnarray}
where $g(k_i,k_j)$, $d(k_i,k_j)$ are
\begin{eqnarray}
g(k_i,k_j) = \frac{2 k_j k_i^2}{k_i^2-k_j^2} ,~~
d(k_i,k_j)=\frac{2 k_i^2 k_j^2(k_i^2+k_j^2)}{(k_i^2-k_j^2)^2} \label{27a}
\end{eqnarray}
and therefore~~$u^{[2]} = -2\left( q_0 +\left(\frac{(\Omega_{1,2})_x}{\Omega_{1,2}}\right)_y\right)$.
~In order to have real expressions, we set $k_2$ as the complex conjugate of $k_1$ which means
\begin{eqnarray}
k_1=A + i B,~~
k_2 = A - i B \label{29b}
\end{eqnarray}
 Using Eqs.~(\ref{29b}) in Eq.~(\ref{26}), we obtain
\begin{eqnarray}
\nonumber
&&\Omega_{1,2} = \left[ (A^2-B^2)x +2q_0 y +\left(3a(6A^2 B^2 -A^4-B^4)
-24 q_0^3\frac{A^2-B^2}{(A^2+B^2)^2}\right) t+\frac{A^2+B^2}{2A}\right]^2 \\
\nonumber
&&+ \left[2A B x+12 A B \left(-a(A^2-B^2)+\frac{4 b q_0^3}{(A^2+B^2)^2}\right)t - \frac{A^2+B^2}{2 B}\right]^2 +(B^2-A^2)\left[\frac{A^2+B^2}{2AB}\right]^2 \label{30}\\
\end{eqnarray}
which for $B^2 > A^2$ has no zeros which means that Eq.~(\ref{30}) does not have singularities. Actually, it is possible to define a Galilean transformation of the following form,
\begin{eqnarray}
\nonumber
x = X + X_0 +v_x t,~~
y = Y +Y_0 +v_y t,~~
X_0 = \frac{A^2+B^2}{4 A B^2},~~
Y_0 = -\frac{(A^2+B^2)^2}{8 q_0 A B^2},\\
\nonumber
v_x = \left[6a(A^2-B^2)-\frac{24 b q_0^3}{(A^2+B^2)^2}\right],~~
v_y = \frac{1}{q_0} \left[-\frac{3a}{2} (A^2+B^2)^2+24 b q_0^3 \frac{(A^2-B^2)}{(A^2+B^2)^2}\right] \label{31}\\
\end{eqnarray}
such that in the new coordinates, $\Omega_{1,2}$ reads as the static solution
\begin{eqnarray}
\Omega_{1,2} = \left[(A^2-B^2) X + 2 q_0 Y \right]^2 +\left[2 A B X\right]^2 +(B^2-A^2)\left[\frac{A^2+B^2}{2AB}\right]^2
\end{eqnarray}
Similarly, one can define  $v^{[2]}$ and $w^{[2]}$. The lump solution for $u^{[2]}$ is represented in Fig. 1. It is interesting to note that one gets a similar lump profile for $v^{[2]}$ and $w^{[2]}$.
\begin{figure}
\centerline{\includegraphics[height = 52.0 mm,width = 52.0 mm]{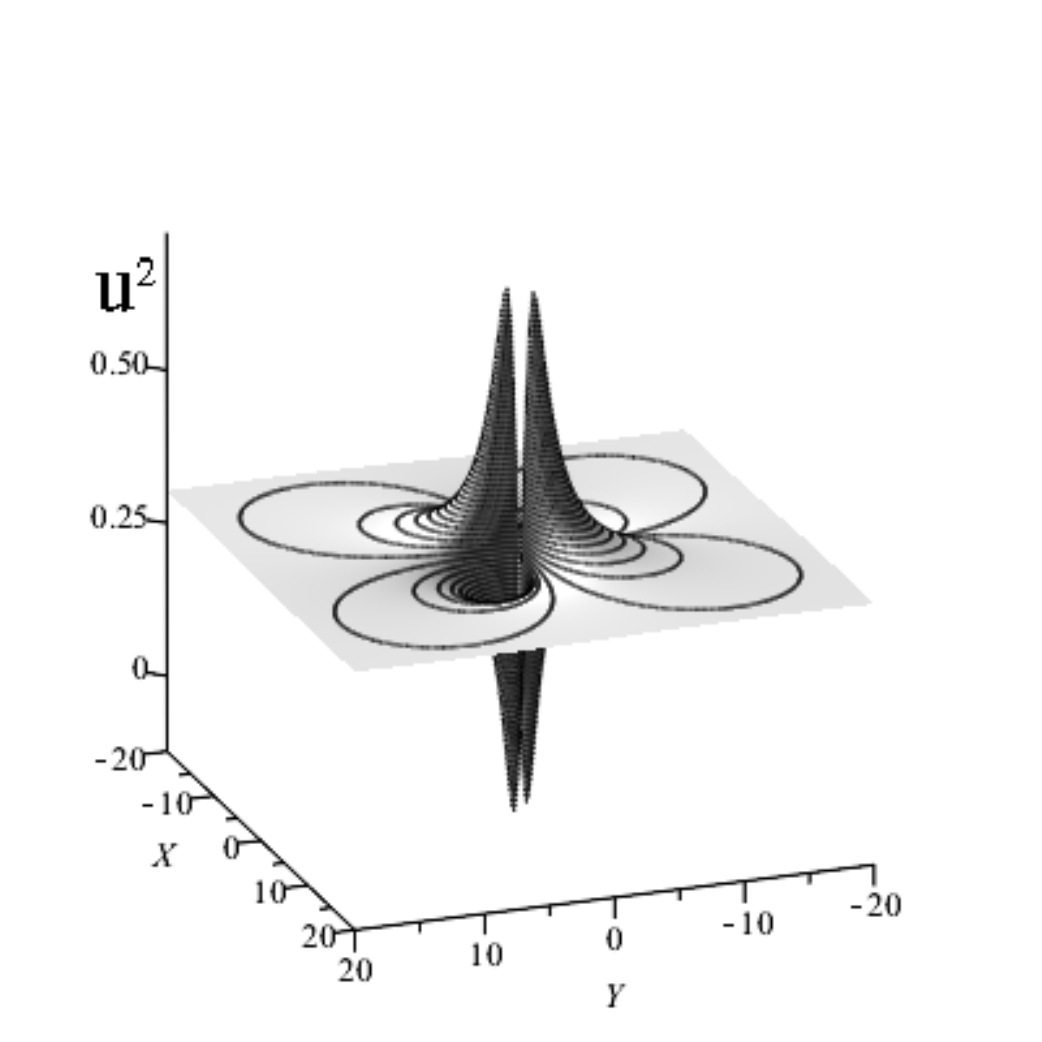}}
\caption{Lump for $n=1$ when $q_0 =0.3$, $A=0.5$, B=1}
\end{figure}
\subsection{Case-II: $n=2$}
Substituting  $n=2$ into Eqs.~(\ref{21a})-(\ref{21d}), we have
\begin{eqnarray}
\nonumber
&&\phi_{i}(k_{i}) = exp^{k_{i} (x+J(k_{i}))} \left(\psi(k_{i})^2+ a_0 (k_{i})\right) \label{33a},~~
J(k_{i}) = \frac{-2 q_0 y}{k^2_{i}} +\left(-a k^2_{i} +\frac{8 b q_0^3}{k^4_{i}}\right) t \label{33b},\\
&&\psi(k_{i}) =k^2_{i} \left( x+\frac{2 q_0 y}{k^2_{i}} -3\left( a k^2_{i} +\frac{8 b q^3_0}{k^4_i}\right) t\right) \label{33c}
\end{eqnarray}
From Eqs.~(\ref{22a})-(\ref{22b}),
\begin{equation}
a_0 (k_i) =-4 q_0 k_i y -\frac{6}{k_i} \left( a k_i^6 -16 b q_0^3 \right) t
\end{equation}
We can calculate the matrix $\Delta_{i,j}$ through the integration of Eq.~(\ref{15}) as,
\begin{eqnarray}
\nonumber
&&\Delta_{i,j} \left( \frac{k_1+k_2}{2 k_1}\right) exp^{     -k_i(x+J(k_i))-k_j(x+J(k_j))}=\\
\nonumber
&&\psi(k_1)^2 \psi(k_2)^2
-\frac{2 k_2^2 }{k_1+k_2}\psi(k_1)^2 \psi(k_2) 
+\frac{2 k_1 k_2}{k_1 +k_2} \psi(k_1) \psi (k_2)^2
+\left( a_0(k_2) +\frac{2 k_2^4}{(k_1+k_2)^2}\right)\psi(k_1)^2\\
\nonumber
&&+\left( a_0(k_1)
-\frac{2 k_2 k_1^3}{(k_1+k_2)^2}\right) \psi(k_2)^2
+\frac{2 k_1 k_2}{k_1+k_2} \left( a_0(k_2)
+ \frac{2 k_2^3 (k_2-2 k_1)}{(k_1+k_2)^2}\right)
 \psi(k_1)
- \frac{2 k_2^2}{k_1+k_2}\left( a_0(k_1)
 +\frac{2 k_1^3 (k_1 -2 k_2)}{(k_1+k_2)^2}\right)\\
&&\psi(k_2)
 + a_0(k_1) a_0(k_2)
+\frac{2 k_2}{(k_1+k_2)^2}\left( k_2^3 a_0(k_1)- k_1^3 a_0(k_2)\right)
+\frac{4 k_2^2k_1 (k_1-k_2)}{(k_1+k_2)^2} \psi(k_1) \psi(k_2)+\frac{12 k_1^3 k_2^4(k_1-k_2)}{(k_1+k_2)^4}
\end{eqnarray}
\subsection*{$\tau - function$}
A second iteration provides,~~ $m^{[2]} = q_0 x y + \ln(\tau_{1,2})$
\begin{eqnarray}
&&u^{[2]} =-2(m_{x,y}^{[2]})=-2\left( q_0 +\left(\frac{(\tau_{1,2})_x}{\tau_{1,2}}\right)_y \right)
\end{eqnarray}
where, $\tau_{1,2}  = \phi_1 \phi_2 -\Delta_{1,2} = \frac{1}{2}   \left(\Delta_{2,1} -\Delta_{1,2}\right)$, which after simplification can be written as,
\begin{eqnarray}
\nonumber
&&\tau_{1,2} = -\frac{k_1 - k_2}{k_1 +k_2} exp^{k_{1} (x+J(k_{1}))+k_2(x+J(k_{2}))} \Omega_{1,2},\\
\nonumber
&&\Omega_{1,2} =\left[ (\psi(k_1)+g(k_1,k_2)) ^2 +a_0(k_1)-\frac{k_1}{k_2} g(k_1,k_2)^2\right] 
\left[(\psi(k_2)+g(k_2,k_1))^2
+a_0(k_2)-\frac{k_2}{k_1} g(k_2,k_1)^2)\right] \\
&&+4 d(k_1,k_2)\left[(\psi(k_1)-c(k_1,k_2))(\psi(k_2)-c(k_2,k_1))\right]
+p(k_1,k_2)
\end{eqnarray}
where $g(k_i,k_j)$, $d(k_i,k_j)$ are defined in Eq.~(\ref{27a}) and
\begin{eqnarray}
\nonumber
c(k_i,k_j) =k_i^2 \frac{k_i^2 -k_i k_j +2 k_j^2}{(k_i+k_j) (k_i^2+k_j^2)},~~
p(k_i,k_j) = \frac{8 k_i^4 k_j^4 (k_i^2+k_j^2+k_ik_j)}{(k_i^2+k_j^2) (k_i+k_j)^4}
\end{eqnarray}
Since $c(k_i,k_j)$, $p(k_i,k_j)$ are constants, we have ~~$u^{[2]} = -2\left( q_0 +\left(\frac{(\Omega_{1,2})_x}{\Omega_{1,2}}\right)_y\right)$. If we select $k_1=A + i B$, $k_2 = A- i B$, we obtain  the real expression for $\Omega_{1,2}$ as
\begin{eqnarray}
\nonumber
&&\Omega_{1,2} = \left[ \left(A^2 -B^2)X +2 q_0 Y\right)^2 - 4 A^2 B^2 X^2  - 4 A q_0 Y
+8 A^2 B^2 h_1 t +\frac{(3 A^2 -B^2)(A^4-B^4)}{4 A^2 B^2} \right]^2\\
\nonumber
&&+\left[4(A^2-B^2)A B X^2 +8 q_0 A B X Y
-4 q_0 B Y
-8 A^2 B^2 h_2 t
+\frac{(3 A^2 -B^2)(A^2+B^2)}{2 A B} \right] ^2\\ 
\nonumber
&&+(B^2-A^2)
\left[\frac{A^2+B^2}{2AB}\right]^2
\left[(A^2-B^2)X
+2 q_0 Y -\frac{2 A^4 -A^2 B^2 -B^4}{2A(A^2-B^2)}\right]^2\\
&&+(B^2-A^2) \left[\frac{A^2+B^2}{2AB}\right]^2  \left[2 A B X +\frac{A^4- A^2 B^2 +2 B^4}{2 B (A^2 -B^2)}\right]^2 
+ \frac{B^2-3 A^2}{B^2 - A^2}\left[\frac{(A^2+B^2)^2}{2 A^2}\right]^2 \label{41}
\end{eqnarray}
where $h_1$ and $h_2$ are  constants defined by
\begin{eqnarray}
\nonumber
h_1 =\frac{3}{A(A^2 + B^2)^2}\left[8 b q_0^3 +a \left(3 A^6 -B^6 +5 A^4 B^2 +A^2  B^4 \right) \right],\\
h_2 = \frac{3}{B(A^2+B^2)} \left[8 b q_0^3 -a\left(3 B^6 -A^6 +5 B^4 A^2 +B^2 A^4\right) \right] \label{42b}
\end{eqnarray}
and $X, Y$ are the coordinates defined in Eq.~(\ref{31}).
From Eq.~(\ref{41}), it is easy to see that $\Omega_{1,2}$ does not have zeros when $B^2 > 3 A^2$.
If we wish to study the behavior of the solution when $t\rightarrow \pm\infty$, we need to perform the transformation, ~~$X = X_{\infty} \pm \sqrt{c t}$, $Y = Y_{\infty} \pm z  \sqrt{c t}$
and fix $c$ and $z$ to cancel the higher powers in $t$ of Eq.~(\ref{41}). The result is
\begin{eqnarray}
\nonumber
c^2 - 2 h_1 c -h_2^2=0 \Rightarrow c=h_1\pm\sqrt{h_1^2+h_2^2},~~
z=\frac{B^2-A^2}{2 q_0}+ \frac{A B h_2}{q_0 c}
\end{eqnarray}
In this case, at $t\rightarrow \pm\infty$, $\Omega_{1,2}$ behaves as
\begin{eqnarray}
\nonumber
&&\Omega_{1,2} \sim \left[\left(2 h_2 (A^2-B^2) -4 A B c\right)X_\infty
+4 q_0 h_2 Y_\infty -2 A h_2 +\left(\frac{A^2-B^2}{B}\right)c\right]^2\\
\nonumber
&&\left[\left(4 A B h_2+2(A^2-B^2)c\right)X_\infty+4 q_0 c Y_\infty
-2 B h_2 +\left(\frac{A^2-B^2}{A}\right)c\right]^2\\
&&+(h_2^2+c^2)(B^2-A^2)\left[\frac{A^2+B^2}{2AB}\right]^2 \label{45}
\end{eqnarray}
which corresponds to a static lump. Let us consider the two possible solutions of Eq.~(\ref{45}) separately.\\
$\bullet$  At $t\longrightarrow -\infty$
\begin{eqnarray}
\nonumber
c_- = -\sqrt{h_1^2+h_2^2} +h_1 < 0\Rightarrow c_-t>0,~~
z_- = \frac{B^2 - A^2}{2 q_0} -\frac{A B h_2}{q_0  (\sqrt{h_1^2+h_2^2}+h_1)}
\end{eqnarray}
There are two lumps approaching along the line,~~$X = X_{-\infty} \pm \sqrt{c_- t}$,~~$Y = Y_{-\infty} \pm z_-\sqrt{c_- t}$,~~$Y-Y_{-\infty} = \tan(\theta_-) (X-X_{-\infty})$
\begin{eqnarray}
\nonumber
&&tg(\theta_-) = z_- = \frac{B^2-A^2}{2 q_0 }-\frac{A B h_2}{q_0 \left(-\sqrt{h_1^2 +h_2^2}+h_1\right)}
\end{eqnarray}
$\bullet$ At $t\longrightarrow \infty$
\begin{eqnarray}
\nonumber
c_+ = \sqrt{h_1^2+h_2^2} +h_1 > 0\Rightarrow c_+t>0
\end{eqnarray}
There are again two lumps moving away along the line,~~$X = X_{+\infty} \pm \sqrt{c_+ t}$, ~$Y = Y_{+\infty} \pm z_+\sqrt{c_+ t}$,~ $Y-Y_{+\infty} = \tan(\theta_+) (X-X_{+\infty})$
and therefore,
\begin{eqnarray}
\nonumber
&&\tan(\theta_+) = z_+ = \frac{B^2-A^2}{2 q_0} -\frac{A B h_2}{q_0 \left(\sqrt{h_1^2 +h_2^2}+h_1\right)}
\end{eqnarray} 
The scattering angle between the lumps is given by,
\begin{eqnarray}
\nonumber
\tan (\theta) = \tan(\theta_+ -\theta_-)=  \frac{8 q_0 A B \sqrt{h_1^2+h_2^2}}{4 q_0^2 h_2 +4 A B h_1 (A^2 -B^2) +h_2 (A^4 -6 A^2 B^2 +B^4)}
\end{eqnarray}
\begin{figure}[ht]
\centerline{a) $t<0$\includegraphics[height = 32.0 mm,width = 32.0 mm]{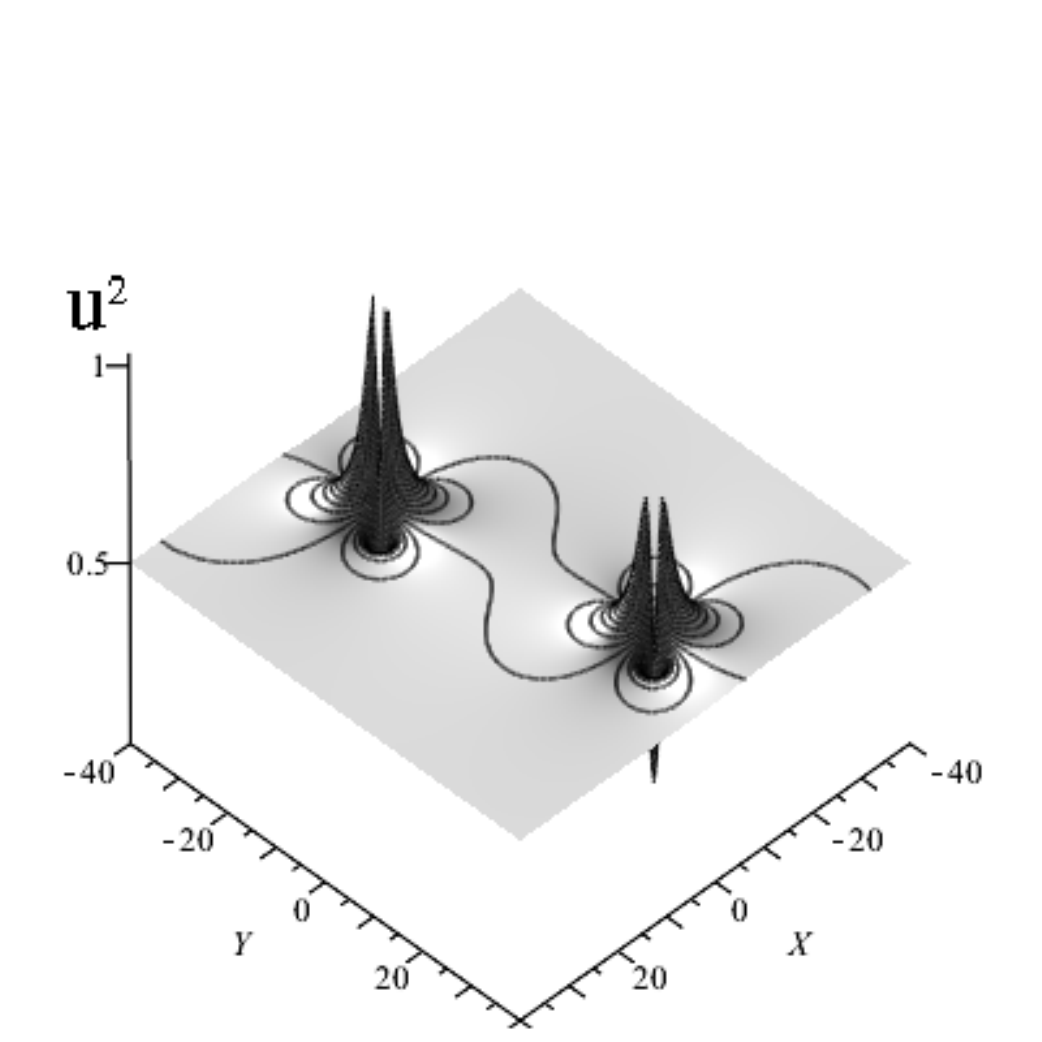}
b) $t=0$\includegraphics[height = 32.0 mm,width = 32.0 mm]{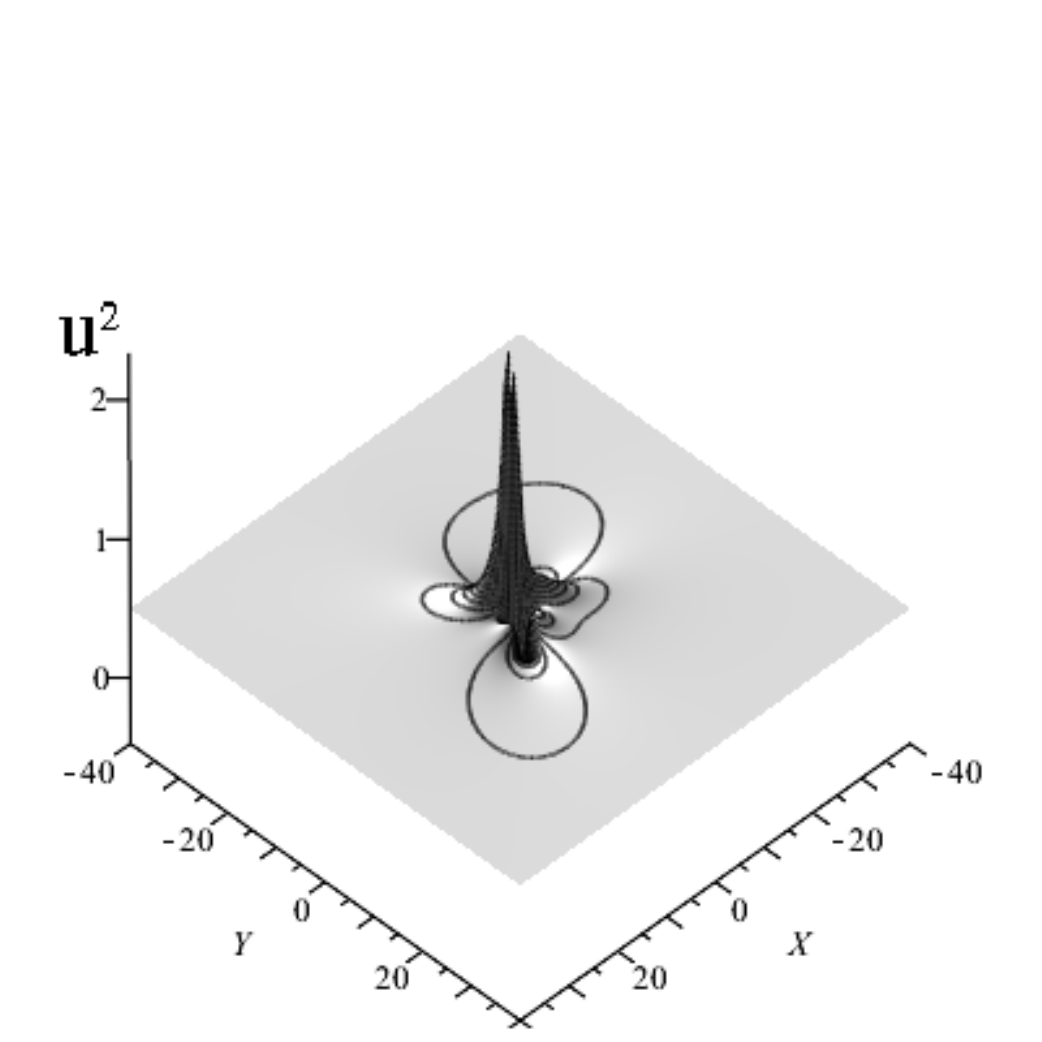}
c) $t>0$\includegraphics[height = 32.0 mm,width = 32.0 mm]{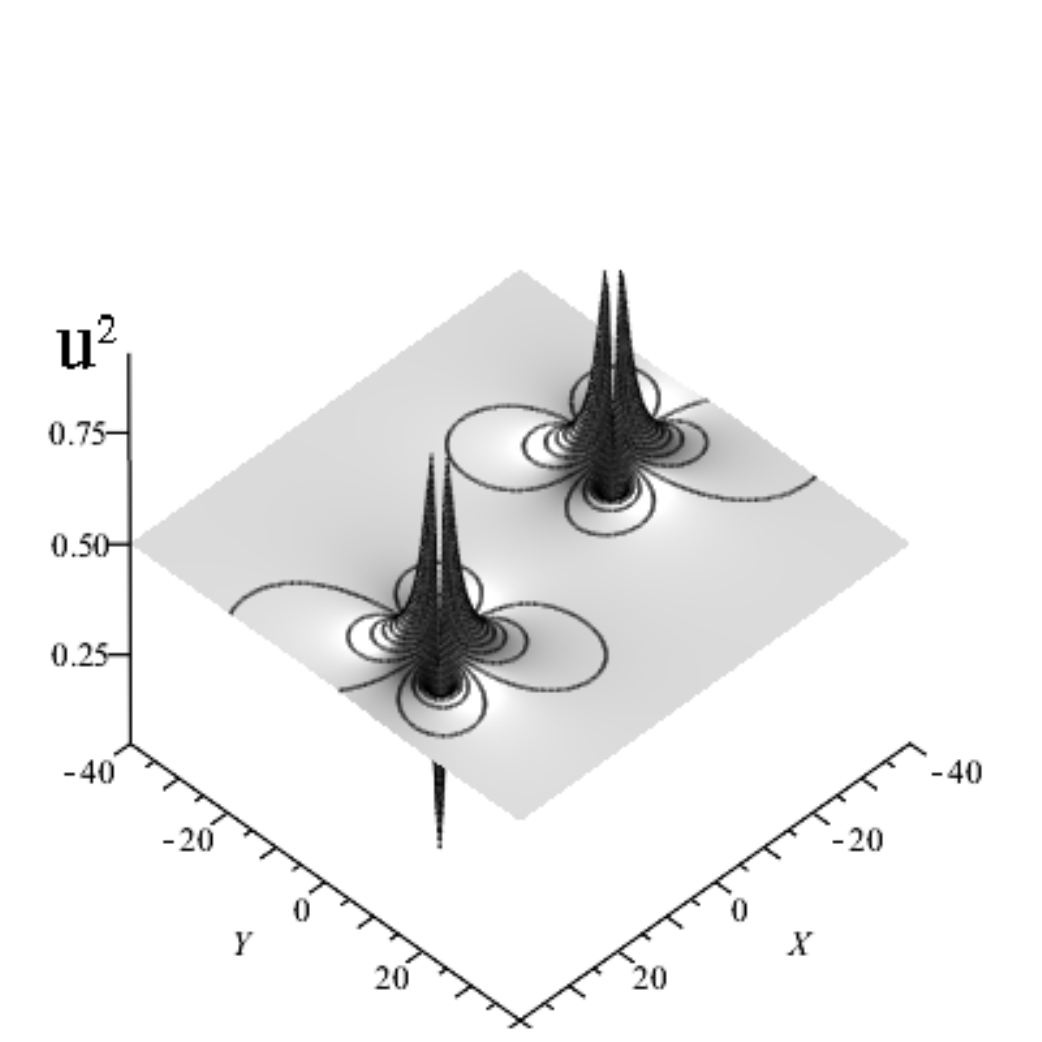}}
\caption  { Lump for $n=2$, when $q_0=0.5$, $a=1$, $b=66$, $A=0.5$, $B=1$}
\label{fig}
\end{figure}
Similarly, one can define  $v^{[2]}$ and $w^{[2]}$.  The lump solution for $u^{[2]}$ is shown in Fig. 2. It is again interesting to note that one gets the same lump profile for $v^{[2]}$ and $w^{[2]}$. From Fig. 2, one understands that there is only a rotation of lumps without any interaction (or exchange of energy). Fig. 2(b) shows the coalesced state of two lump solution, wherein the two lumps just pass through each other.
\subsection{Two lump solution}
As we have seen in the previous section, the one lump solution is obtained through the second iteration. It obviously means that for the two lump solution, we need to go to the fourth iteration. If we start with the singular manifold $\phi_1$, we can generalize Eq.~(\ref{13}) and Eq.~(\ref{16}) as:
\begin{eqnarray}
\nonumber
\phi_{1,j} =\phi_j -\frac{\Delta_{1,j}}{\phi_{1}},~~
\Delta_{i,j} = \Delta(\phi_i,\phi_j)
\end{eqnarray}
From the fourth iteration, we have
\begin{eqnarray}
\nonumber
\phi_{1,i,j,k} = \phi_{1,i,k} - \frac{\Delta_{1,i,j,k}}{\phi_{1,i,j}},~~
\nonumber
\Delta_{1,i,j.k} = \Delta(\phi_{1,i,j} , \phi_{1,i,k}) =\Delta_{i,j,k}- \frac{\Delta_{1,j,i} \Delta_{1,i,k}}{\phi_{1,i}^2}
\end{eqnarray} 
The solution becomes
\begin{eqnarray}
\nonumber
&&m^{[4]} = m^{[3]}+\ln(\phi_{1,2,3,4})
 = m^{[2]} +\ln(\phi_{1,2,3}) +\ln (\phi_{1,2,3,4}) \\
 \nonumber
&&= m^{[1]}
+\ln (\phi_{1,2})+\ln (\phi_{1,2,3})+\ln (\phi_{1,2,3,4})\\
\nonumber
&& = m^{[0]} +\ln (\phi_{1})+\ln (\phi_{1,2}) +\ln (\phi_{1,2,3})
 +\ln (\phi_{1,2,3,4})
\end{eqnarray}
which reads
\begin{eqnarray}
m^{[4]} =m^{[0]}+ \ln(\tau_{1,2,3,4})
\end{eqnarray}
where,
$\tau_{1,2,3,4} = \phi_{1,2,3,4} \phi_{1,2, 3} \phi_{1,2} \phi_1$.
With the previous definition, we can construct the $\tau$ function for the fourth iteration from the  eigenfunctions of the seed solution $m^{[0]}$ in the following form:
\begin{eqnarray}
\nonumber
&&\tau_{1,2,3,4} = \frac{1}{4} (\Delta_{2,1} -\Delta_{1,2})(\Delta_{4,3} - \Delta_{3,4})
-\frac{1}{4} (\Delta_{4,2} -\Delta_{2,4})(\Delta_{3,1} - \Delta_{1,3}) \\
\nonumber
&& + \frac{1}{4} (\Delta_{4,1} -\Delta_{1,4})(\Delta_{3,2} - \Delta_{2,3})
\end{eqnarray}
where we have used, $\phi_i\phi_j =\frac{1}{2} (\Delta_{j,i}+\Delta_{i,j})$.
One can write  $\tau_{1,2,3,4}$ in a more compact form as:
$\tau^2_{1,2,3,4} = det(\Delta_{i,j}), \,  \textrm{if}  \,\,i,j = 1..4$. We shall consider the simplest case in which we have the seed solutions with $n=1$.
\subsubsection{Solution for two lumps with $n=1$}
The simplest case can be obtained by taking $n=1$. The eigenfunction given by Eqs.~(\ref{21a})-(\ref{21d}) again taking the form given by Eq.(\ref{23a})-(\ref{23b}). We can  calculate the matrix $\Delta_{i,j}$ again taking the form given by Eq.~(\ref{24}). We have,

\begin{eqnarray}
u^{[2]} = -2\left( q_0 +\left(\frac{(\tau_{1,2,3,4})_x}{\tau_{1,2,3,4}}\right)_y\right)
\end{eqnarray}
and we choose
\begin{eqnarray}
\nonumber
k_1 = A_1 +i B_1,~~
k_2 = k_1^* = A_1 - i B_1,~~
k_3 =  A_2 + i B_2,~~
k_4 = k_3^* = A_2- i B_2
\end{eqnarray}
It is convenient to define a center of mass coordinate system as
\begin{eqnarray}
x = X_{cm} +\frac{1}{2} (v_x^1 + v_x^2) t,~~
y = Y_{cm} +\frac{1}{2} (v_y^1+ v_y^2) t \label{59}
\end{eqnarray}
where $(v_x^i,v_y^i)$ are the individual velocities of each soliton (see Eq.~(\ref{31}))
\begin{eqnarray}
\nonumber
v_x^i = \left(6a(A^2_i-B^2_i)-\frac{24 b q_0^3}{(A^2_i+B^2_i)^2}\right),~~
v_y^i= \frac{1}{q_0} \left(\frac{-3a}{2} \left( A^2_i+B^2_i\right)^2+24 b q_0^3 \frac{(A^2_i-B^2_i)}{(A^2_i+B^2_i)^2}\right) \label{60}\\
\end{eqnarray}
Using the change of variables given in Eqs.~(\ref{59})-(\ref{60}) in Eq.~(\ref{23b}), we have
\begin{eqnarray}
\nonumber
\psi(k_1) = k_1^2 (X_{cm} -V_x t)+2q_0 (Y_{cm} -V_y t),~~
\psi(k_2) = k_2^2 (X_{cm} -V_x t)+2q_0 (Y_{cm} -V_y t),\\
\nonumber
\psi(k_3) = k_3^2 (X_{cm} +V_x t)+2q_0 (Y_{cm} +V_y t),~~
\psi(k_4) = k_4^2 (X_{cm} +V_x t)+2q_0 (Y_{cm} +V_y t)
\end{eqnarray}
where, $V_x = \frac{1}{2} (v_x^1 -v_x^2),~~
V_y= \frac{1}{2} (v_y^1 -v_y^2)$.~~
In the center of mass system, the solution asymptotically yields two lumps that move with equal and opposite velocities. To clarify this point, we can consider the asymptotic behavior of each lump\\
$\bullet$ Let  us define
\begin{eqnarray}
\nonumber
X_{cm} = X_1 - X_{0}^1 +V_x t,~~
Y_{cm} = Y_1 - Y_{0}^1 +V_y t
\end{eqnarray}
which  (the tedious calculation has been made with MAPLE) allows us to write the limit of the $\tau$-function when $t\rightarrow \pm\infty$ as the static lump
\begin{eqnarray}
\nonumber
\tau_{1,2,3,4} \sim  \left[(A_1^2-B_1^2) X_1 + 2 q_0 Y_1 \right]^2 +\left[2 A_1 B_1 X_1\right]^2 + (B_1^2-A_1^2)\left[\frac{
 (A_1^2+B_1^2)}{2 A_1 B_1}\right]^2
\end{eqnarray}
where
\begin{eqnarray}
\nonumber
&&X_0^1 =-\frac{A_1^2+B_1^2}{4 A_1 B_1^2}\\
\nonumber
&&-\frac{4 A_2 \left[(A^2_2+B^2_2)^3+(A^2_2-3B^2_2)(A_1^2+B_1^2)^2+2(A^2_2+B^2_2)^2(B_1^2-A^2_1)\right]}{[(A_1+A_2)^2+(B_1-B_2)^2][(A_1-A_2)^2+(B_1-B_2)^2][(A_1+A_2)^2+(B_1+B_2)^2][(A_1-A_2)^2+(B_1+B_2)^2]}
\end{eqnarray}
\begin{eqnarray}
\nonumber
&&q_0Y_0^1=\frac{A_1^2+B_1^2}{8 A_1 B_1^2} \\
\nonumber
&& -\frac{2 A_2(A_1^2+B_1^2)^2[(A^2_1+B^2_1)^2+(A^2_2-3B^2_2)(A^2_2+B^2_2) +2 (A_2^2+B^2_2) (B_1^2-A_1^2)]}{[(A_1+A_2)^2+(B_1-B_2)^2][(A_1-A_2)^2+(B_1-B_2)^2][(A_1+A_2)^2+(B_1+B_2)^2][(A_1-A_2)^2+(B_1+B_2)^2]}
\end{eqnarray}
$\bullet$ If we now define
\begin{eqnarray}
\nonumber
X_{cm} = X_2 - X_{0}^2 -V_x t,~~
Y_{cm} = Y_2 - Y_{0}^2 -V_y t
\end{eqnarray}
the limit of the $\tau$-function when $t\rightarrow \pm\infty$ as the static lump becomes
\begin{eqnarray}
\nonumber
\tau_{1,2,3,4} \sim  \left[(A_2^2-B_2^2) X_2 + 2 q_0 Y_2\right]^2 +\left[2 A_2 B_2 X_2\right]^2 + (B_2^2-A_2^2)\left[\frac{(A_2^2+B_2^2)}{2 A_2 B_2}\right]^2
\end{eqnarray}
where
\begin{eqnarray}
\nonumber
&&X_0^2= -\frac{A_2^2+B_2^2}{4 A_2 B_2^2} \\
\nonumber
&&- \frac{4 A_1 \left[(A_1^2+B_1^2)^3+(A_1^2-3B_1^2)(A_2^2+B_2^2)^2+2(A_1^2+B_1^2)^2(B_2^2-A_2^2)\right]}{[(A_1+A_2)^2+(B_1-B_2)^2][(A_1-A_2)^2+(B_1-B_2)^2][(A_1+A_2)^2+(B_1+B_2)^2][(A_1-A_2)^2+(B_1+B_2)^2]}
\end{eqnarray}
\begin{eqnarray}
\nonumber
&&q_0Y_0^2= \frac{A_2^2+B_2^2}{8 A_2 B_2^2}\\
\nonumber
&& -\frac{2 A_1(A_2^2+B_2^2)^2[(A^2_2+B^2_2)^2+(A^2_1-3B^2_1)(A^2_1+B^2_1)
+2 (A_1^2+B_1^2) (B_2^2-A_2^2)]}{[(A_1+A_2)^2+(B_1-B_2)^2][(A_1-A_2)^2+(B_1-B_2)^2][(A_1+A_2)^2+(B_1+B_2)^2][(A_1-A_2)^2+(B_1+B_2)^2]}
\end{eqnarray}
In this system of reference, the asymptotic behaviour  of the solution for $t \longrightarrow \pm \infty $ corresponds to two lumps moving with equal and opposite velocities along parallel lines as shown in Fig. (3a) and Fig (3c), Fig. (3b) again represents the coalesced state of two lump solution where again the lumps which seem to merge move away in opposite directions later. Similarly, one can define  $v^{[2]}$ and $w^{[2]}$.
\begin{figure}[ht]
\centerline{a) $t<0$\includegraphics[height = 34.0 mm,width = 34.0 mm]{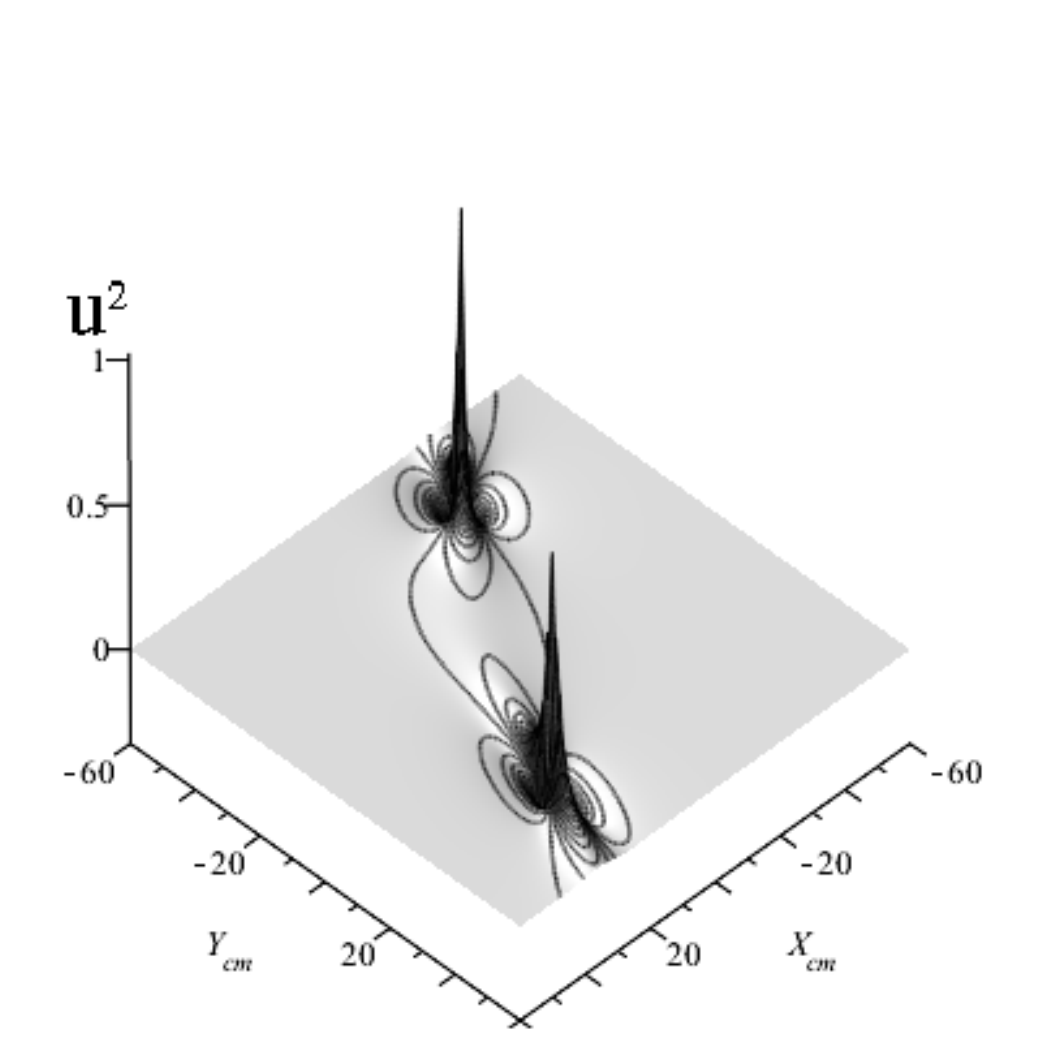}
b) $t=0$\includegraphics[height = 34.0 mm,width = 34.0 mm]{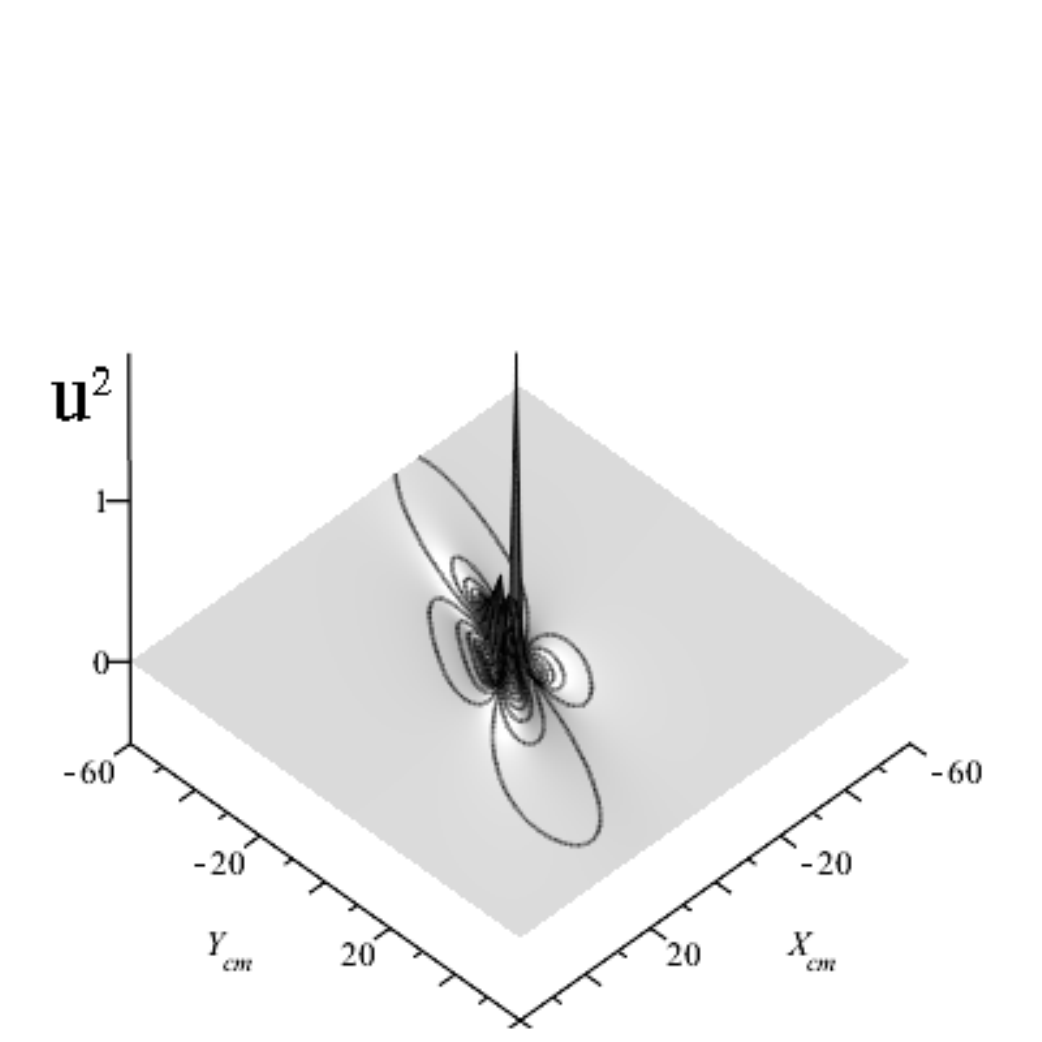}
c) $t>0$\includegraphics[height = 34.0 mm,width = 34.0 mm]{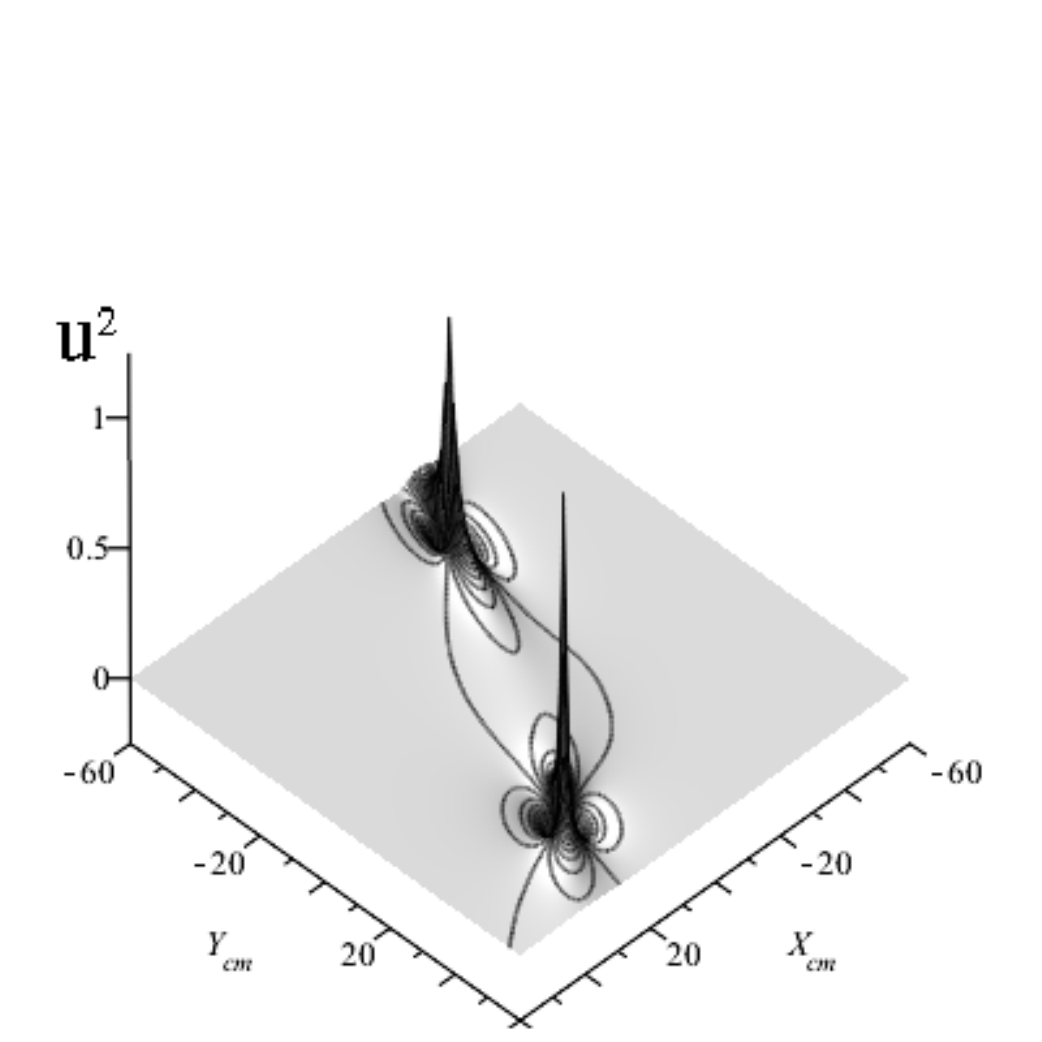}}
\caption  {Two Lump solution for $n=1$,  when $a=1$, $b=0.2$, $q_0=0.5$, $A_1=0.5$, $B_1=1$, $A_2=0.5$, $B_2=\frac{4}{3}$}
\label{fig}
\end{figure}
\section{Rogue waves}
In the section, we will focus on the construction of rogue waves for Eq.~(\ref{3}).
\subsection{Solution}
Taking the easiest choice of the variable $m(x,y,t)$ as,
\begin{eqnarray}
m = A(x,t) +B(y,t) \label{61}
\end{eqnarray}
where $A$ and $B$ are arbitrary functions in the indicated variables, we now substitute equation Eq.~(\ref{61}) in Eq.~(\ref{2a}) to obtain
\begin{eqnarray}
u^{[0]} = -2 m^{[0]}_{xy} = 0\\
v^{[0]} = \frac{c}{3 a} - 2 m^{[0]}_{xx} = \frac{c}{3a} -2 A_{xx} = v(x,t)\\
w^{[0]} = \frac{d}{3b} -2 m^{[0]}_{yy} = \frac{d}{3b} - 2 B_{yy} =w(y,t)
\end{eqnarray}
One possibility is to choose
\begin{eqnarray}
\phi_1 = F(x,t),~ \phi_2 = G(y,t) \label{62}
\end{eqnarray}
where $F(x,t)$ and $G(y,t)$ are again arbitrary functions. Substituting equations Eq.~(\ref{61}) and Eq.~(\ref{62}) in Eqs.~(\ref{11a})-(\ref{11b}), we have
\begin{eqnarray}
\nonumber
F_t + G_t =-a\left(F_{xxx} + 6 F_{x} A_{xx}\right)-b\left(G_{yyy} +6 G_y B_{yy}\right)
\end{eqnarray}
where
\begin{eqnarray}
\nonumber
A_{xx} = - \frac{F_t + a F_{xxx}}{6 a F_x},~~
B_{yy} = -\frac{G_t + b G_{yyy}}{6 b G_y}
\end{eqnarray}
From  Eq.~(\ref{16}), we have ~$\Delta_{1,2} = 2 F(x,t) G(y,t)+c_0$,~
where $c_0$ is an arbitrary constant.
Hence, Eq.~(\ref{18}) now yields
\begin{eqnarray}
\nonumber
\tau_{1,2} = - \left( F(x,t) G(y,t) + c_0 \right)
\end{eqnarray}
Now, the solution for $u^{[2]}$, $v^{[2]}$ and $w^{[2]}$, can be written as,
\begin{eqnarray}
\nonumber
m^{[2]} = m^{[0]} + \ln(\tau_{1,2}) \label{63a}\\
\nonumber
u^{[2]} = - 2 m^{[2]}_{xy} = -2 \left(m^{[0]}+\ln \left(-\left( F(x,t) G(y,t) +c_0 \right) \right) \right)_{xy} \label{63b}\\
\nonumber
v^{[2]} = \frac{c}{3 a} - 2 m^{[2]}_{xx} =\frac{c}{3 a} -2 \left(m^{[0]}+\ln \left(- \left(F(x,t) G(y,t) +c_0 \right) \right) \right)_{xx} \label{63c}\\
\nonumber
w^{[2]} = \frac{d}{3 b} - 2 m^{[2]}_{yy} = \frac{d}{3 b}-2 \left(m^{[0]}+\ln \left( \left(- (F(x,t) G(y,t) +c_0\right) \right) \right)_{yy} \label{63d}
\end{eqnarray}
where
\begin{eqnarray}
\nonumber
F = f(x,t) + \frac{c_3}{c_4},~~
G= g(y,t) +\frac{c_2}{c_4},~~
c_0 =  \frac{c_1}{c_4} - \frac{c_2 c_3}{c_4^2} \label{64a}
\end{eqnarray}
\subsection{Case-I}
To construct a single rogue wave, we choose
\begin{eqnarray}
\nonumber
f(x,t) = \frac{1}{1+t^2 +(x-1)^2},~~
g(y,t) = 2 y^2
\end{eqnarray}
Rogue waves for $u^{[2]}$, $v^{[2]}$ and $w^{[2]}$ are shown in Fig. 4. The time evolution of the rogue waves indicates their unstable nature.
\begin{figure}[ht]
\centerline{a)\includegraphics[height = 42.0 mm,width = 42.0 mm]{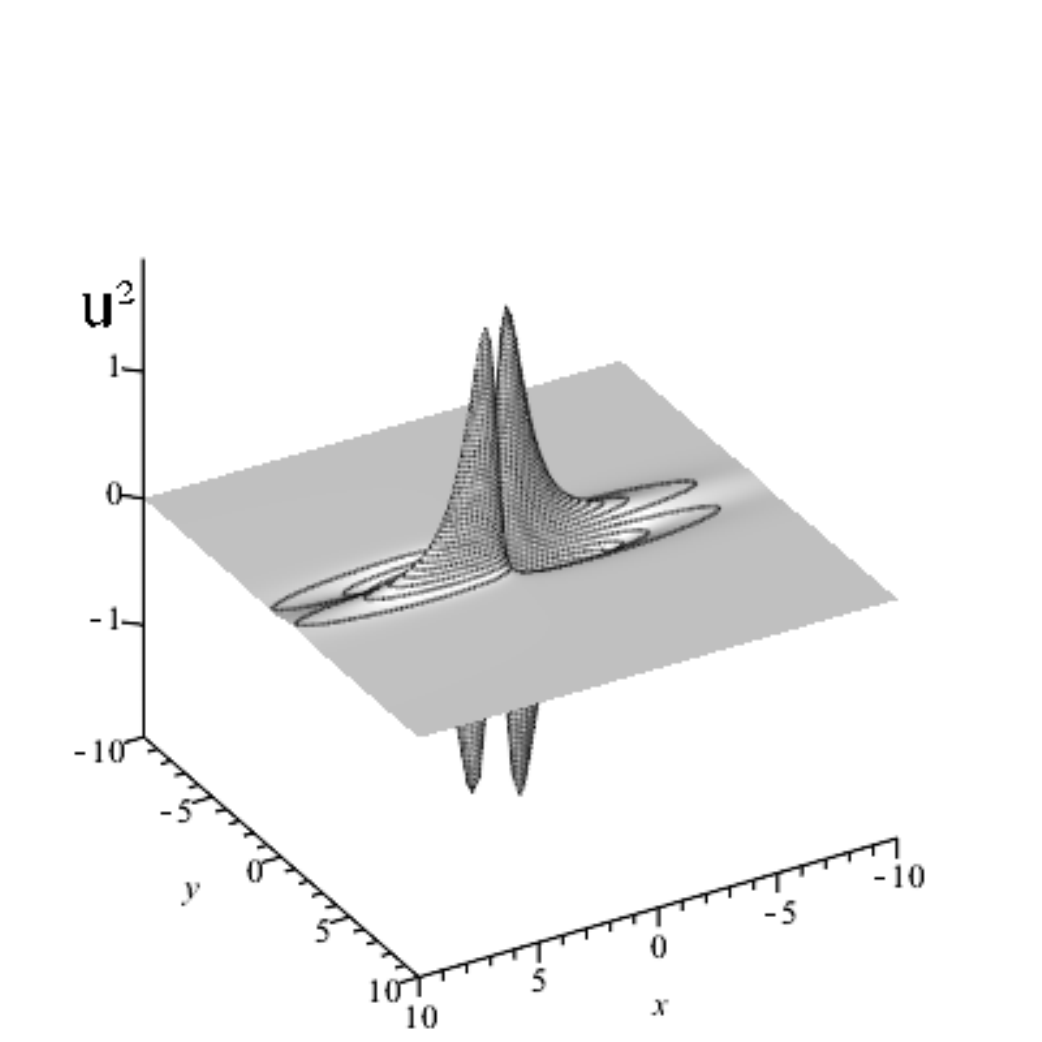}
b) \includegraphics[height = 42.0 mm,width = 42.0 mm]{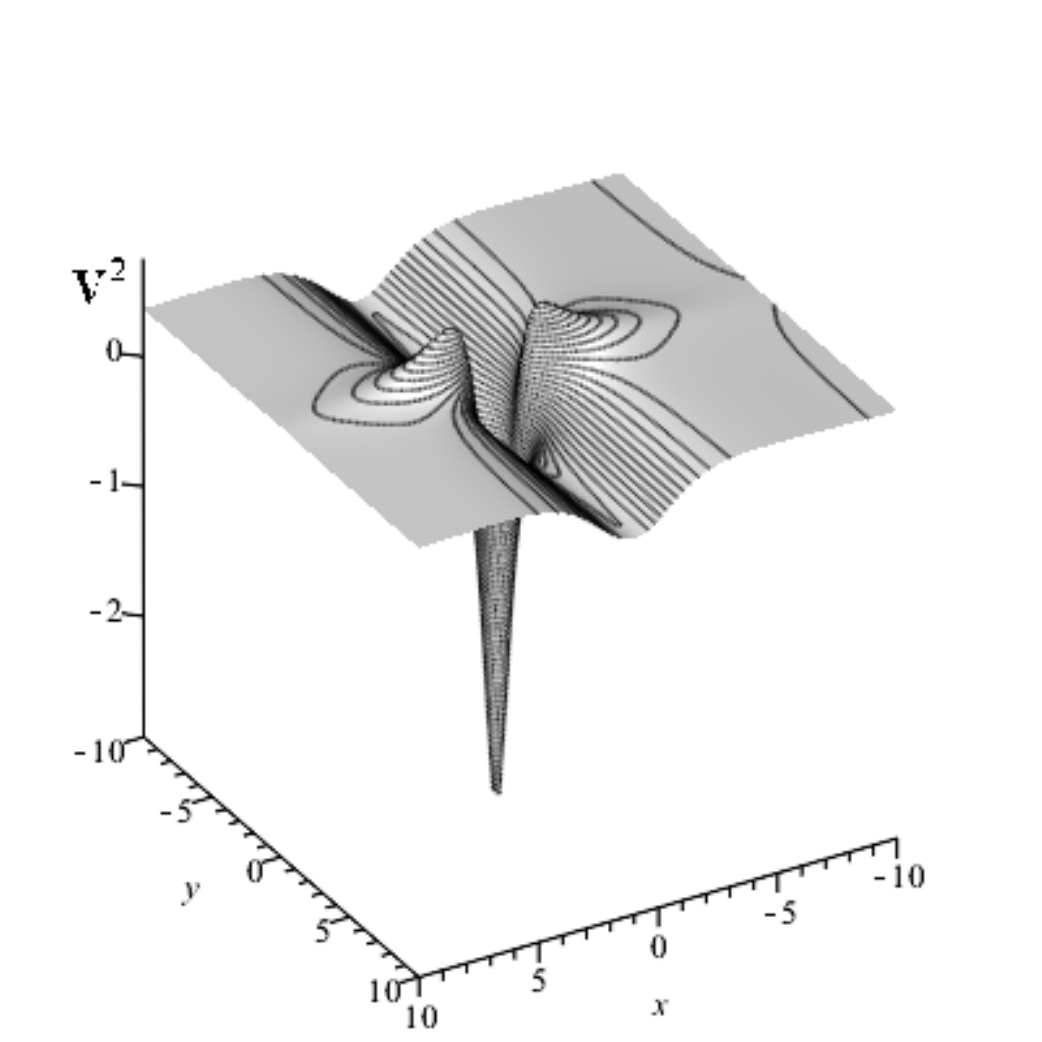}
c)\includegraphics[height = 42.0 mm,width = 42.0 mm]{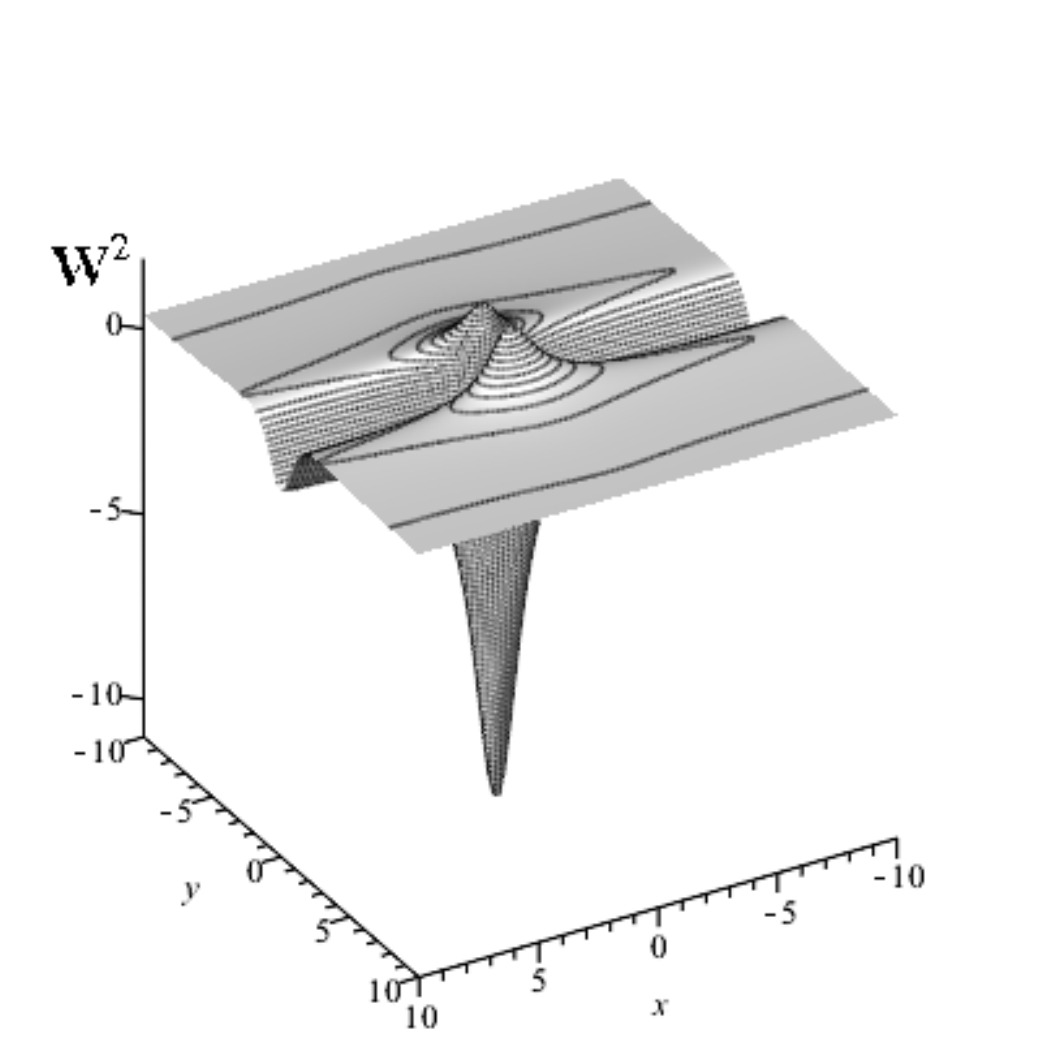}}
\caption {Rogue waves for $u^{[2]}$,  $v^{[2]}$ and $w^{[2]}$ with  $c_1 = 0.06$,~ $c_2 = 0.01$, ~$c_3 = 0.01$,~ $c_4 = 0.1$, ~$a =1,~ b = 1, ~c =1,~ d = 1 $ at $t=0$}
\label{fig }
\end{figure}
\subsection{Case-II}
To obtain  a multi rogue waves, we choose,
\begin{eqnarray}
\nonumber
f(x,t) = \cos (x)+ \sin(x),~~
g(y,t) = \frac{1}{(1+(y-1)^2+ k  t^2)^2}
\end{eqnarray}
Multi rogue waves for $u^{[2]}$, $v^{[2]}$ and $w^{[2]}$ are shown in Fig. 5.
\begin{figure}[ht]
\centerline{a)\includegraphics[height = 42.0 mm,width = 42.0 mm]{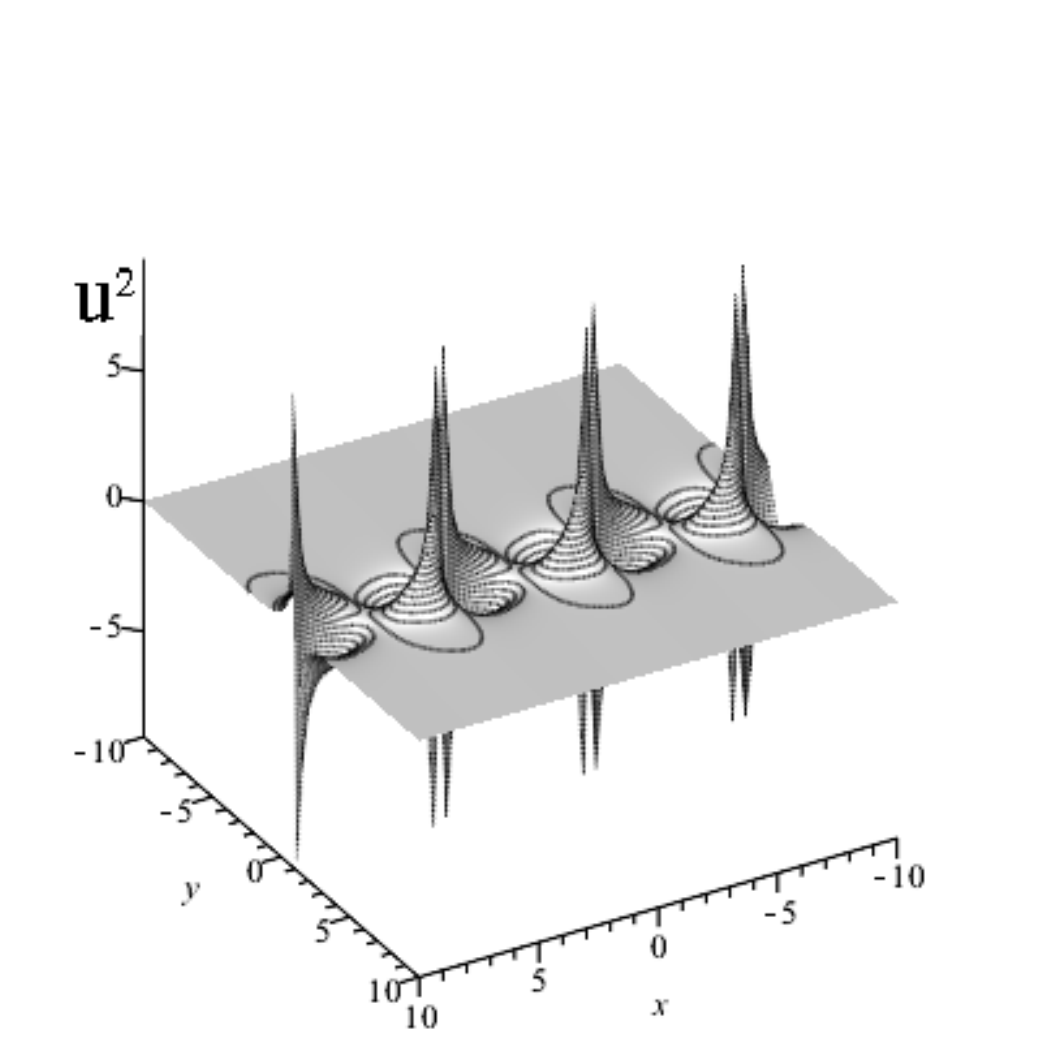}
b) \includegraphics[height = 42.0 mm,width = 42.0 mm]{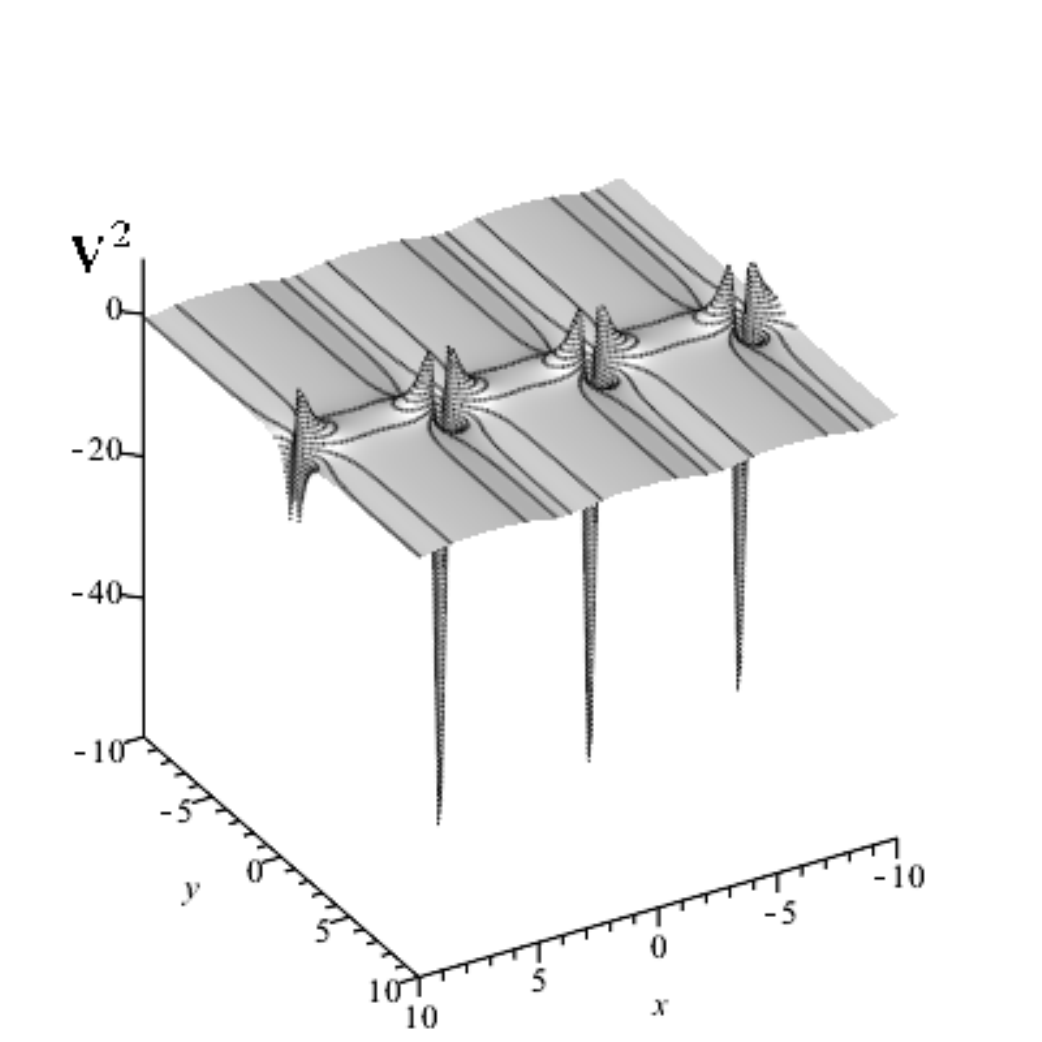}
c) \includegraphics[height = 42.0 mm,width = 42.0 mm]{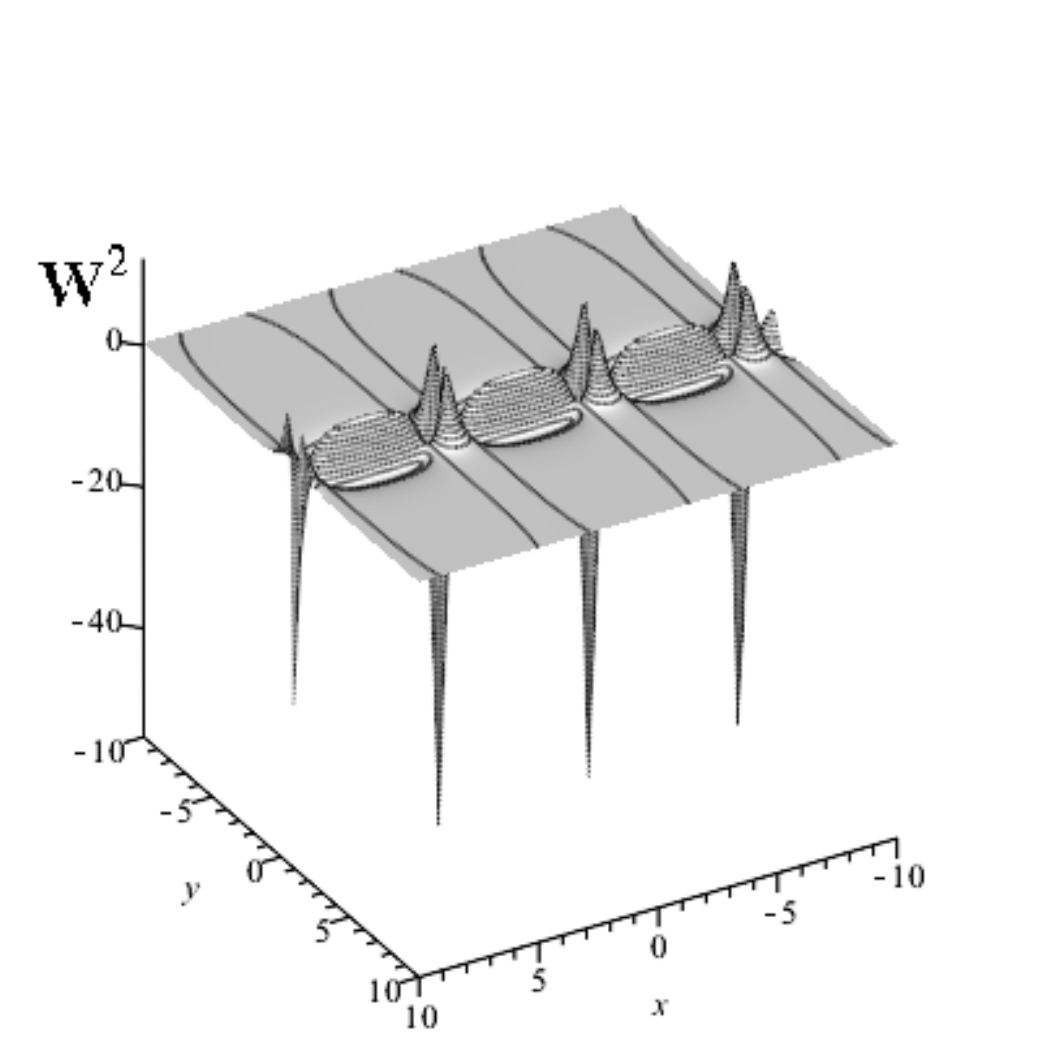}}
\caption {Multi rogue waves for $u^{[2]}$, $v^{[2]}$ and $w^{[2]}$ with $c_1 = 0.06$,~ $c_2 = 0.01$,~ $c_3 = 0.1$,~ $c_4 = 0.1$,~ $a =15,~ b = 10, ~c =10, ~d = 10, ~$k= 0.01$ $ at $t=0$}
\label{fig}
\end{figure}
\section{Discussion}
In this paper, we have analyzed the generalized NNV equation (GNNV) and derived its Lax-pair in the coordinate space destroying the myth of weak Lax-pair. We have then generated lumps and rogue waves of the GNNV equation and studied their dynamics. The lumps do not interact and they merely pass through each other  or move away from each other while the rogue waves generated are found to retain their unstable nature. We believe that a deeper investigation may unearth other elusive localized solutions.

\section{Acknowledgements}
R.S wishes to thank Department of Atomic Energy -
National Board of Higher Mathematics (DAE-NBHM) for providing a Junior Research Fellowship.  R.R. acknowledges DST (grant No. SR/S2/HEP-26/2012),
Council of Scientific and Industrial Research (CSIR), India (grant
03(1323)/14/EMR-II dated 03.11.2014) and Department of Atomic Energy -
National Board of Higher Mathematics (DAE-NBHM), India (grant 2/48(21)/2014/NBHM(R.P.) /R $\&$ D II/15451 ) for financial support in the form of Major Research Projects. The research of P. G. E has been supported in part by MINECO (project MAT2013-46308 and MAT2016-75955) and Junta de Castillay Le\'on (project SA045U16).

\end{document}